\documentclass[11pt]{article}
\usepackage{moriond,epsfig}
\usepackage[latin1]{inputenc}
\usepackage[OT1]{fontenc}
\usepackage{bm}
\usepackage{amsmath}
\usepackage{amsfonts}

\def\be{\begin{equation}}
\def\ee{\end{equation}}
\def\bea{\begin{eqnarray}}
\def\eea{\end{eqnarray}}

 \newcommand{\ud}{\mathrm{d}}
 \newcommand{\calW}{\mathcal{W}}

\begin{document}
\vspace*{4cm} \title{PHENOMENOLOGY OF THE MODIFIED NEWTONIAN DYNAMICS \\AND
THE CONCORDANCE COSMOLOGICAL SCENARIO}

\author{L. BLANCHET \& A. LE TIEC}

\address{$\mathcal{G}\mathbb{R}\varepsilon{\mathbb{C}}\mathcal{O}$,
Institut d'Astrophysique de Paris --- UMR 7095 du CNRS, \\ Universit\'e Pierre
\& Marie Curie, 98\textsuperscript{bis} boulevard Arago, 75014 Paris, France}

\maketitle\abstracts{After reviewing the modified Newtonian dynamics (MOND)
  proposal, we advocate that the associated phenomenology may actually not
  result from a modification of Newtonian gravity, but from a mechanism of
  ``gravitational polarization'' of some dipolar medium playing the role of
  dark matter. We then build a relativistic model within standard general
  relativity to describe (at some phenomenological level) the dipolar dark
  matter polarizable in a gravitational field. The model naturally involves a
  cosmological constant, and is shown to reduce to the concordance
  cosmological scenario ($\Lambda$-CDM) at early cosmological times. From the
  mechanism of gravitational polarization, we recover the phenomenology of
  MOND in a typical galaxy at low redshift. Furthermore, we show that the
  cosmological constant $\Lambda$ scales like $a_0^2$, where $a_0$ is the
  constant MOND acceleration scale, in good agreement with observations.}

\section{Introduction}

The mysteries of the nature of dark matter and dark energy are perhaps the
most important ones of contemporary cosmology. Dark matter, which accounts for
the observed discrepancy between the dynamical and luminous masses of bounded
astrophysical systems, is usually formulated within the so-called
\textit{particle dark matter} approach, in which the dark matter consists of
unknown non-baryonic particles, e.g. neutralinos as predicted by
super-symmetric extensions of the standard model of particle physics
(see~\cite{Be.al.05} for a review). Furthermore, the dark matter triggers the
formation of large-scale structures by gravitational collapse and explains the
distribution of baryonic matter from galaxy cluster scales up to cosmological
scales by the non-linear growth of initial perturbations. Simulations suggest
some universal dark matter density profile around distributions of ordinary
baryonic matter~\cite{Na.al.97}. An important characteristic of dark matter,
required by the necessity of clustering matter on small scales, is that it
should be cold (or non-relativistic) at the epoch of galaxy
formation. Together with the observational evidence of dark energy (presumably
a cosmological constant $\Lambda$) measured from the Hubble diagram of
supernovas, the particle dark matter hypothesis has yielded the successful
concordance model of cosmology called $\Lambda$-CDM, which reproduces
extremely well the observed cosmic microwave background spectrum~\cite{HuD02}.

However, despite these successes at cosmological scales, the particle dark
matter approach has some difficulties~\cite{McBl.98} at explaining in a
natural way the flat rotation curves of galaxies, one of the most persuasive
evidence for the existence of dark matter, and the Tully-Fisher empirical
relation between the observed luminosity and the asymptotic rotation velocity
of spiral galaxies. In order to deal with these difficulties, all linked with
the properties of dark matter at galactic scales, an alternative paradigm has
emerged in the name of the modified Newtonian dynamics
(MOND)~\cite{Mi1.83,Mi2.83,Mi3.83}. Although MOND in its original formulation
cannot be considered as a viable physical model, it is now generally admitted
that it does capture in a very simple and powerful ``phenomenological recipe''
a large number of observational facts, that any pertinent model of dark matter
should explain.

It is frustrating that the two alternatives $\Lambda$-CDM and MOND, which are
successful in complementary domains of validity (say the cosmological scale
for $\Lambda$-CDM and the galactic scale for MOND), seem to be fundamentally
incompatible. In the present paper, we shall propose a different approach,
together with a new interpretation of the phenomenology of MOND, which has the
potential of bringing together $\Lambda$-CDM and MOND into a single unifying
relativistic model for dark matter and dark energy. This relativistic model
will be shown to benefit from both the successes of $\Lambda$-CDM at
cosmological scales, and MOND at galactic scales.

\section{The modified Newtonian dynamics (MOND)}

The original idea behind MOND~\cite{Mi1.83,Mi2.83,Mi3.83} is that there is no
dark matter, and we witness a violation of the fundamental law of gravity (or
of inertia). MOND is designed to account for the basic features of galactic
dark matter halos. It states that the ``true'' gravitational field $\bm{g}$
experienced by ordinary matter, say a test particle whose acceleration would
thus be $\bm{a}=\bm{g}$, is not the Newtonian gravitational field
$\bm{g}_\mathrm{N}$, but is actually related to it by
\begin{equation}\label{recipe1}
\mu\!\left(\frac{g}{a_0}\right)\bm{g} = \bm{g}_\mathrm{N} \, .
\end{equation}
Here $\mu$ is a function of the dimensionless ratio $g/a_0$ between the norm
of the gravitational field $g=\vert\bm{g}\vert$, and the constant MOND
acceleration scale $a_0 \simeq 1.2 \times 10^{-10}~\text{m} / \text{s}^2$,
whose numerical value is chosen to fit the data. The specific MOND regime
corresponds to the weak gravity limit, much weaker than the scale $a_0$. In
this regime (where formally $g \rightarrow 0$) we
have~\cite{Mi1.83,Mi2.83,Mi3.83}
\begin{equation}\label{recipe2}
\mu\!\left(\frac{g}{a_0}\right) = \frac{g}{a_0} + \mathcal{O}(g^2) \, .
\end{equation}
On the other hand, when $g$ is much larger than $a_0$ (formally $g \rightarrow
+\infty$) the usual Newtonian law is recovered, i.e. $\mu \rightarrow
1$. Various functions $\mu$ interpolating between the MOND regime and the
Newtonian limit are possible, but most of them appear to be rather \textit{ad
hoc}. Taken for granted, the MOND ``recipe''~\eqref{recipe1}--\eqref{recipe2}
beautifully predicts a Tully-Fisher relation and is very successful at fitting
the detailed shape of rotation curves of galaxies from the observed
distribution of stars and gas (see~\cite{Mi.02,SaMc.02} for reviews). So MOND
appears to be more than a simple recipe and may well be related to some new
fundamental physics. In any case the agreement
of~\eqref{recipe1}--\eqref{recipe2} with a large number of observations calls
for a clear physical explanation.

Taking the divergence of both sides of~\eqref{recipe1}, and using the usual
Poisson equation for the Newtonian field $\bm{g}_\mathrm{N}$, we obtain a
local formulation of MOND in the form of the modified Poisson equation
\cite{BeMi.84}
\begin{equation}\label{MONDeq}
\bm{\nabla} \! \cdot \! \left( \mu \, \bm{g} \right) = -4 \pi G \,
	\rho_\text{b} \, ,
\end{equation}
where $\rho_\text{b}$ is the density of baryonic matter. This equation can be
derived from a Lagrangian, and that Lagrangian has been the starting point for
constructing relativistic extensions of MOND. Such extensions postulate the
existence of extra (supposedly fundamental) fields associated with gravity
besides the spin-$2$ field of general relativity. Promoting the Newtonian
potential $U$ to a scalar field $\phi$, scalar-tensor theories for MOND have
been constructed~\cite{BekS94} but shown to be non viable; essentially because
light signals do not feel the presence of the scalar field, since the physical
(Jordan-frame) metric is conformally related to the Einstein-frame metric, and
the Maxwell equations are conformally invariant. This is contrary to
observations: huge amounts of dark matter are indeed observed by gravitational
(weak and strong) lensing.

Relativistic extensions of MOND that pass the problem of light deflection by
galaxy clusters have been shown to require the existence of a time-like vector
field. The prototype of such theories is the tensor-vector-scalar (TeVeS)
theory~\cite{Sa.97,Be.04,Sa.05}, whose non-relativistic limit reproduces MOND,
and which has been extensively investigated in cosmology and at the
intermediate scale of galaxy clusters~\cite{An.al.06}. Modified gravity
theories such as TeVeS have evolved recently toward Einstein-{\ae}ther like
theories~\cite{JaMa.01,Zl.al.07,Ha.al.08}. Still, recovering the level of
agreement of the $\Lambda$-CDM scenario with observations at cosmological
scales remains an issue for such theories. In the present paper we shall
follow a completely different route from that of modified gravity and/or
Einstein-{\ae}ther theories. We shall propose an alternative to these theories
in the form of a specific \textit{modified matter theory} based on an
elementary interpretation of the MOND equation~\eqref{MONDeq}.

\section{Interpretation of the phenomenology of MOND}

The physical motivation behind our approach is the striking (and presumably
deep) analogy between MOND and the electrostatics of dielectric
media~\cite{Bl1.07}. From electromagnetism in dielectric media we know that
the Maxwell-Gauss equation can be written in the two equivalent forms
\begin{equation}
	\bm{\nabla} \! \cdot \! \bm{E} = \frac{1}{\varepsilon_0} \, \bigl(
	\sigma_\text{free} + \sigma_\text{pol} \bigr) \quad
	\Longleftrightarrow \quad \bm{\nabla} \! \cdot \! \left(
	\varepsilon_\text{r} \, \bm{E} \right) = \frac{1}{\varepsilon_0} \,
	\sigma_\text{free} \, ,
\end{equation}
where $\bm{E}$ is the electric field, $\sigma_\text{free}$ and
$\sigma_\text{pol}$ are the densities of free and polarized (electric) charges
respectively, and $\varepsilon_\text{r} = 1 + \chi_\text{e}$ is the relative
permittivity of the dielectric medium. Such an equivalence is only possible
because the density of polarized charges reads $\sigma_\text{pol} = -
\bm{\nabla}\!\cdot\!\bm{P}$, where the polarization field $\bm{P}$ is
aligned with the electric field according to $\bm{P} = \varepsilon_0
\chi_\text{e} \, \bm{E}$, the proportionality coefficient $\chi_\text{e}(E)$
being known as the electric susceptibility. 

By full analogy, one can write the MOND equation \eqref{MONDeq} in the form of
the usual Poisson equation but sourced by some additional distribution of
``polarized gravitational masses'' $\rho_\text{pol}$ (to be interpreted as
dark matter, or a component of dark matter), namely
\begin{equation}\label{Poisson_MOND}
	\bm{\nabla} \! \cdot \bm{g} = -4 \pi G \, \bigl( \rho_\text{b} +
	\rho_\text{pol} \bigr) \quad \Longleftrightarrow \quad
	\bm{\nabla} \! \cdot \! \left( \mu \, \bm{g} \right) = -4 \pi G \,
	\rho_\text{b} \, .
\end{equation}
This rewriting stands as long as the mass density of polarized masses appears
as the divergence of a vector fied, namely takes the \textit{dipolar} form
\begin{equation}\label{rhodm}
	\rho_\text{pol} = - \bm{\nabla}\!\cdot\!\bm{\Pi}\, ,
\end{equation}
where $\bm{\Pi}$ denotes the (gravitational analogue of the) polarization
field. It is aligned with the gravitational field $\bm{g}$ (i.e. the dipolar
medium is polarized) according to
\begin{equation}\label{Pig}
	\bm{\Pi} = -\frac{\chi}{4 \pi G} \, \bm{g} \, .
\end{equation}
Here the coefficient $\chi$, which depends on the norm of the gravitational
field $g=\vert\bm{g}\vert$ in complete analogy with the electrostatics of
dielectric media, is related to the MOND function by
\begin{equation}\label{muchi}
	\mu=1+\chi\, .
\end{equation}
Obviously $\chi$ can be interpreted as a ``gravitational susceptibility''
coefficient, while $\mu$ itself can rightly be called a ``digravitational''
coefficient. It was shown in~\cite{Bl1.07} that in the gravitational case the
sign of $\chi$ should be negative, in perfect agreement with what MOND
predicts; indeed, we have $\mu<1$ in a straightforward interpolation between
the MOND and Newtonian regimes, hence $\chi<0$. This finding is in contrast to
electromagnetism where $\chi_\text{e}$ is positive. It can be viewed as some
``anti-screening'' of gravitational (baryonic) masses by polarization masses
--- the opposite effect of the usual screening of electric (free) charges by
polarization charges. Such anti-screening mechanism results in an enhancement
of the gravitational field \textit{\`a la} MOND, and offers a very nice
interpretation of the MOND phenomenology. Furthermore, it was pointed out that
the stability of the dipolar dark matter medium requires the existence of some
internal force, which turned out to be simply (in a crude quasi-Newtonian
model~\cite{Bl1.07}) that of an harmonic oscillator. This force could then be
interpreted as the restoring force in the gravitational analogue of a plasma
oscillating at its natural plasma frequency. Finally, it seems from this
discussion that the dielectric interpretation of MOND is deeper than a mere
formal analogy. However the model~\cite{Bl1.07} is clearly non-viable because
it is non-relativistic, and it involves negative gravitational-type masses and
therefore a violation of the equivalence principle at a fundamental level.

\section{Relativistic model for the dipolar dark fluid}\label{sec3}

Here we shall take seriously the physical intuition that MOND has something to
do with a mechanism of gravitational polarization. We shall build a fully
relativistic model based on a matter action in standard general
relativity. Note that this means we are changing the point of view of the
original MOND proposal. Instead of requiring a modification of the laws of
gravity in the absence of dark matter, we advocate that the phenomenology of
MOND results from a physically well-motivated mechanism acting on a new type
of dark matter, very exotic compared to standard particle dark matter. Thus,
we are proposing a modification of the \textit{dark matter} sector rather than
a modification of gravity as in TeVeS like theories.

\subsection{Action and equations of motion}

From the previous discussion, the necessity of endowing dark matter with a new
vector field to build the polarization field is clear. However this vector
field will not be expected to be fundamental as in TeVeS like
theories. Extending previous work~\cite{Bl2.07}, our model will be based on a
\textit{matter} action (in Eulerian fluid formalism) in general relativity of
the form
\begin{equation}\label{S}
	S = \int \ud^4 x \, \sqrt{-g} \, L \bigl[ J^\mu, \xi^\mu,
	\dot{\xi}^\mu, g_{\mu \nu} \bigr] \, .
\end{equation}
This action is to be added to the Einstein-Hilbert action for gravity, and to
the actions of all the other matter fields. It contains \textit{two} dynamical
variables: (i) a conserved current $J^\mu = \sigma u^\mu$ satisfying
$\nabla_\mu J^\mu=0$, where $u^\mu$ is the normalized four-velocity and
$\sigma = (- J_\nu J^\nu)^{1/2}$ is the rest mass energy density (we pose
$c=1$ throughout); (ii) the vector field $\xi^\mu$ representing the dipole
four-vector moment carried by the fluid particles. This extra field being
dynamical, the Lagrangian will also depend on its covariant derivative
$\nabla_\nu \xi^\mu$; but in our model this dependence will occur only through
the covariant time derivative $\dot{\xi}^\mu \equiv u^\nu \nabla_\nu
\xi^\mu$. The Lagrangian explicitly reads (see~\cite{BlLe.08} for details)
\begin{equation}\label{L}
	L = \sigma \left[ -1 - \Xi + \frac{1}{2} \dot{\xi}_\mu \dot{\xi}^\mu
	\right] - \calW (\Pi_\perp) \, ,
\end{equation}
where $\Xi \equiv \bigl\{ \bigl( u_\mu - \dot{\xi}_\mu \bigr) \bigl( u^\mu -
\dot{\xi}^\mu \bigr) \bigr\}^{1/2}$. The first term is a mass term in the
ordinary sense (i.e. the Lagrangian of a pressureless perfect fluid), the
second one is inspired by the action of spinning particles in general
relativity~\cite{BaIs.80}, and the third term is a kinetic term for the dipole
moment. Finally, the last term represents a potential $\calW$ describing some
internal interaction, function of the polarization (scalar) field $\Pi_\perp =
(\perp_{\mu\nu}\!\Pi^\mu\Pi^\nu)^{1/2}$, where $\Pi^\mu = \sigma \xi^\mu$ is
the polarization four-vector, and $\perp_{\mu\nu} \, = g_{\mu\nu} + u_\mu
u_\nu$ is the projector orthogonal to the four-velocity.

By varying the action \eqref{S}--\eqref{L} with respect to both the current
$J^\mu$ and the dipole moment $\xi^\mu$, we get two dynamical equations: an
equation of motion for the dipolar fluid, and an evolution equation for the
dipole moment $\xi^\mu$. From these equations it can be shown~\cite{BlLe.08}
that we can impose the constraint $\Xi = 1$ as a particular way of selecting a
physically interesting solution, such that the final equations depend only on
the \textit{space-like} projection $\xi_\perp^\mu=\,\perp_{\mu\nu}\!\xi^\nu$
of the dipole moment $\xi^\mu$. The final equations we obtain are
\begin{align}
\dot{u}^\mu &= - \mathcal{F}^\mu \equiv - \hat{\xi}_\perp^\mu \, \calW' \, ,
\label{motion} \\ \dot{\Omega}^\mu &= \frac{1}{\sigma} \nabla^\mu \left( \calW
- \Pi_\perp \calW' \right) - \xi_\perp^\nu R^\mu_{\phantom{\mu} \rho \nu
\lambda} u^\rho u^\lambda \, , \label{evolution}
\end{align}
where we denote $\Omega^\mu \equiv u^\mu \bigl( 1 + \xi_\perp \calW' \bigr) \,
+ \!  \perp^\mu_\nu \dot{\xi}_\perp^\nu$, and employ the notations
$\hat{\xi}_\perp^\mu \equiv \xi_\perp^\mu / \xi_\perp = \Pi_\perp^\mu /
\Pi_\perp$ and $\calW' \equiv \ud \calW / \ud \Pi_\perp$. The motion of the
dipolar fluid as given by \eqref{motion} is non-geodesic, and driven by the
internal force $\mathcal{F}^\mu$ derived from the potential $\calW$. Observe
the coupling to the Riemann curvature tensor in the equation of evolution
\eqref{evolution} of the dipole moment. By varying the action with respect to
the metric $g_{\mu \nu}$ we obtain the stress-energy tensor $T^{\mu
\nu}$. Using the canonical decomposition $T^{\mu \nu} = r \, u^\mu u^\nu +
\mathcal{P} \! \perp^{\mu \nu} + \, 2 \, Q^{(\mu} u^{\nu)} + \Sigma^{\mu
\nu}$, we find the energy density $r$, pressure $\mathcal{P}$, heat flux
$Q^\mu$ (such that $u_\mu Q^\mu=0$) and anisotropic stresses $\Sigma^{\mu
\nu}$ ($u_\nu \Sigma^{\mu\nu}=0$ and $\Sigma^\nu_\nu=0$) as
\begin{subequations}\label{rPQSigma}\begin{align}
		r &= \calW - \Pi_\perp \calW' \! + \rho \, , \label{r} \\
		\mathcal{P} &= - \calW + \frac{2}{3} \, \Pi_\perp \calW' \, ,
		\label{P} \\ Q^\mu &= \sigma \, \dot{\xi}_\perp^\mu +
		\Pi_\perp \calW' u^\mu - \Pi_\perp^\lambda \nabla_\lambda
		u^\mu \, , \label{Q} \\ \Sigma^{\mu \nu} &= \biggl(
		\frac{1}{3} \! \perp^{\mu \nu} \!  - \, \hat{\xi}_\perp^{\mu}
		\hat{\xi}_\perp^{\nu}\biggr) \Pi_\perp \calW'
		\label{Sigmamunu} \, .
\end{align}\end{subequations}
Here the contribution $\rho$ to the energy density involves a monopolar term
$\sigma$ and a dipolar term $- \nabla_\lambda \Pi_\perp^\lambda$ which clearly
appears as a relativistic generalisation of \eqref{rhodm}, and will play the
crucial role when recovering MOND:
\begin{equation}\label{rho}
	\rho = \sigma - \nabla_\lambda \Pi_\perp^\lambda\, .
\end{equation}

\subsection{Weak field expansion of the internal potential}

The dipolar fluid dynamics in a given background metric, and its influence on
spacetime are now known; in the following we shall apply this model to
large-scale cosmology and to galactic halos. For both applications we shall
need to consider the model in a regime of \textit{weak gravity}, which will be
either first-order perturbations around a Friedman-Lema\^itre-Robertson-Walker
(FLRW) background in cosmology, or the non-relativistic limit for galaxies. A
crucial assumption we make is that the potential function $\calW$ admits a
minimum when the polarization $\Pi_\perp$ is zero, and can be Taylor-expanded
around that minimum, with coefficients being entirely specified (modulo an
overall factor $G$) by the single surface density scale built from the MOND
acceleration $a_0$,
\begin{equation}\label{Sigma}
	\Sigma \equiv \frac{a_0}{2 \pi G}\, .
\end{equation}
These coefficients in the expansion of $\calW$ when $\Pi_\perp\rightarrow 0$
will be \textit{fine-tuned} in order to recover the relevant physics at
cosmological and galactic scales. Physically, this expansion corresponds to
$\Pi_\perp \ll \Sigma$ and is valid in the weak gravity limit $g\ll
a_0$. Clearly, the minimum of $\calW$ is nothing but a cosmological constant
$\Lambda$, and we find
\begin{equation}\label{W}
	\calW(\Pi_\perp) = \frac{\Lambda}{8 \pi G} + 2 \pi G \, \Pi_\perp^2 +
	\frac{8 \pi G}{3 \Sigma} \, \Pi_\perp^3 + \mathcal{O}(\Pi_\perp^4) \,
	.
\end{equation}
The expansion is thereby determined up to third order inclusively.

Our assumption that the function $\calW$ involves the single fundamental scale
$\Sigma$ implies in particular that the cosmological constant $\Lambda$ should
scale with $G^2\Sigma^2\sim a_0^2$. We thus introduce a dimensionless
parameter $\alpha$ through
\begin{equation}\label{Lambda_a0}
	\alpha \, a_0 = \frac{1}{2\pi}\sqrt{\frac{\Lambda}{3}} \equiv
	a_\Lambda \, .
\end{equation}
This parameter represents a conversion factor between $a_0$ and the natural
acceleration scale $a_\Lambda$ associated with the cosmological constant
$\Lambda$. Posing $x \equiv \Pi_\perp / \Sigma$, we find that \eqref{W} can be
recast in the form $\calW(\Pi_\perp) = 6 \pi G \, \Sigma^2 \, w(x)$, where
\begin{equation}\label{W2}
	w(x) = \alpha^2 \pi^2 + \frac{1}{3} x^2 + \frac{4}{9} x^3 +
	\mathcal{O}(x^4) \, .
\end{equation}
The present model should be considered only as ``effective'' or
phenomenological, in the sense that the weak-gravity expansion \eqref{W} or
\eqref{W2} should come from a more fundamental (presumably quantum) underlying
theory. Therefore, the coefficients in \eqref{W2} should not be given by
exceedingly large or small numbers (like $10^{10}$ or $10^{-10}$), but rather
be numerically of the order of one, up to a factor of, say, ten. Hence, we
expect that $\alpha$ itself should be around one, and we see by
\eqref{Lambda_a0} that the cosmological constant $\Lambda$ should naturally be
numerically of the order of $a_0^2$. Notice though that our model does not
provide a way to compute the exact value of $\alpha$. However, one can say
that the ``cosmic coincidence'' (see e.g.~\cite{Mi.02}) between the values of
$a_0$ and $a_\Lambda$ --- with the measured conversion factor being $\alpha
\simeq 0.8$ --- finds a natural explanation if dark matter is made of a fluid
of polarizable gravitational dipole moments.

\section{Recovering the $\Lambda$-CDM scenario at cosmological scales}\label{sec4}

Consider a small perturbation of a background FLRW metric valid between, say,
the end of the inflationary era and the recombination. The dipolar fluid is
described by its four-velocity $u^\mu = \overline{u}^\mu + \delta u^\mu$,
where $\delta u^\mu$ is a perturbation of the background comoving
four-velocity $\overline{u}^\mu=(1,\bm{0})$, and by its rest mass density
$\sigma = \overline{\sigma} + \delta \sigma$, where $\delta \sigma$ is a
perturbation of the mean cosmological value $\overline{\sigma}$. The crucial
point in our analysis of the dipolar fluid in cosmology is that the background
value of the dipole moment field $\xi_\perp^\mu$ (which is orthogonal to the
four-velocity and therefore is \textit{space-like}) must vanish in order to
preserve the spatial isotropy of the FLRW background. We shall thus write
$\xi_\perp^\mu = \delta \xi_\perp^\mu$, and similarly $\Pi_\perp^\mu = \delta
\Pi_\perp^\mu$ for the polarization.

At first perturbation order, the stress-energy tensor with explicit components
\eqref{rPQSigma} can be naturally recast, using also \eqref{W}, in the form
$T^{\mu \nu} = T_\text{de}^{\mu \nu} + T_\text{dm}^{\mu \nu}$, where the
explicit expressions of the dark energy and dipolar dark matter components are
\begin{align}
	T_\text{de}^{\mu \nu} &= - \frac{\Lambda}{8 \pi G} \, g^{\mu
	\nu} \, , \\ T_\text{dm}^{\mu \nu} &= \rho \, u^\mu u^\nu + 2
	\, Q^{(\mu} u^{\nu)} \label{exprTdm}\, .
\end{align}
At that order, we find that the dipolar dark matter density reduces to $\rho$
given by \eqref{rho}. The heat flux in \eqref{exprTdm} reads as $Q^\mu =
\sigma \, \dot{\xi}_\perp^\mu - \Pi_\perp^\lambda \nabla_\lambda u^\mu$; it is
non zero, however it is perturbative because so are both $\xi_\perp^\mu$ and
$\Pi_\perp^\mu$. The point for our purpose is that at linear perturbation
order $Q^\mu$ can be absorbed into a redefinition of the perturbed
four-velocity of the dipolar fluid. Posing $\delta \widetilde{u}^\mu = \delta
u^\mu + Q^\mu / \overline{\sigma}$ and introducing the effective four-velocity
$\widetilde{u}^\mu = \overline{u}^\mu + \delta \widetilde{u}^\mu$, we find
that at first order the dipolar dark matter fluid is described by the
stress-energy tensor
\begin{equation}
 T_\text{dm}^{\mu \nu} = \rho \, \widetilde{u}^\mu \widetilde{u}^\nu \, ,
\end{equation}
which is that of a perfect fluid with four-velocity $\widetilde{u}^\mu$,
vanishing pressure and energy density \eqref{rho}. In the linear cosmological
regime, the dipolar fluid therefore behaves as standard cold dark matter (a
pressureless fluid) plus standard dark energy (a cosmological
constant). Adjusting the background value $\overline{\sigma}$ so that
$\Omega_\mathrm{dm}\simeq 23\%$, the model is thus consistent with the
standard $\Lambda$-CDM scenario and the cosmological observations of the CMB
fluctuations (see~\cite{BlLe.08} for more details).

\section{Recovering the MOND phenomenology at galactic scales}

Next, we turn to the study of the dipolar dark fluid in a typical galaxy at
low redshift. We have to consider the non-relativistic limit ($c \rightarrow
+\infty$) of the model; we consistently neglect all relativistic terms
$\mathcal{O}(c^{-2})$. It is straightforward to check that in the
non-relativistic limit the equation of motion \eqref{motion} of the dipolar
fluid reduces to
\begin{equation}\label{motion_NR}
	\frac{\ud \bm{v}}{\ud t} = \bm{g} - \hat{\bm{\Pi}}_\perp \calW' \, ,
\end{equation}
where $\bm{g}$ is the \textit{local} gravitational field generated by both the
baryonic matter in the galaxy and the dipolar dark matter. Applying next the
standard minimal coupling to gravity in general relativity, we find that the
gravitational field obeys the Poisson equation
\begin{equation}\label{Poisson}
	\bm{\nabla}\!\cdot \bm{g} = -4 \pi G \, \bigl( \rho_\text{b} + \rho
	\bigr) \, ,
\end{equation}
where $\rho_\text{b}$ and $\rho$ are respectively the baryonic and dipolar
dark matter mass densities. From \eqref{rho} the dipolar dark matter density
reduces in the non-relativistic limit to
\begin{equation}\label{rhosigma}
	\rho = \sigma - \bm{\nabla} \! \cdot \! \bm{\Pi}_\perp \, .
\end{equation}
The first term is a usual monopolar contribution: the rest mass density
$\sigma$ of the fluid, while the second term is the dipolar contribution, and
can be interpreted as coming from the fluid's internal energy. Here
$\bm{\Pi}_\perp$ denotes the spatial components of the polarization.

In order to recover MOND, we need two things: (i) to find a mechanism for the
alignment of the polarization field $\bm{\Pi}_\perp$ with the gravitational
field $\bm{g}$, so that a relation similar to \eqref{Pig} will apply; (ii) to
justify that the rest mass density $\sigma$ of dipole moments in
\eqref{rhosigma} is small with respect to the baryonic density
$\rho_\text{b}$, hence the galaxy will mostly appear as baryonic in MOND fits
of rotation curves. We have proposed in~\cite{BlLe.08} a single mechanism able
to answer positively these two points. We call it the hypothesis of
\textit{weak clustering} of dipolar dark matter; it is an hypothesis because
it has been conjectured but not proved, and should be checked using numerical
simulations. The weak clustering hypothesis is motivated by a solution of the
full set of equations describing the non-relativistic motion of dipolar dark
matter in a typical baryonic galaxy whose mass distribution $\rho_\text{b}$ is
spherically symmetric (see the Appendix in~\cite{BlLe.08}). This particular
solution corresponds to an equilibrium configuration in spherical symmetry,
for which $\bm{v} = \bm{0}$ and $\sigma = \sigma_0(r)$. The dipole moments
remain at rest because the gravitational field $\bm{g}$ is balanced by the
internal force $\bm{\mathcal{F}} = \hat{\bm{\Pi}}_\perp \calW'$; for that
solution the right-hand-side of \eqref{motion_NR} vanishes.

During the cosmological evolution we expect that the dipolar medium will not
cluster much because the internal force may balance part of the local
gravitational field generated by an overdensity. The dipolar dark matter
density contrast in a typical galaxy at low redshift should thus be small, at
least smaller than in the standard $\Lambda$-CDM scenario. Hence $\sigma \ll
\rho_\text{b}$, and we could even envisage that $\sigma$ stays around its mean
cosmological value, $\sigma \sim \overline{\sigma} \ll \rho_\text{b}$. Now,
because of its size and typical time-scale of evolution, a galaxy is almost
unaffected by the cosmological expansion of the Universe. The cosmological
mass density $\overline{\sigma}$ of the dipolar dark matter is not only
homogeneous, but also almost constant in this galaxy. The continuity equation
reduces to $\bm{\nabla} \! \cdot \! \left( \overline{\sigma} \, \bm{v} \right)
\simeq \bm{0}$, and the most simple solution corresponds to a static fluid
verifying $\bm{v} \simeq \bm{0}$. By \eqref{motion_NR} we see that the
polarization field $\bm{\Pi}_\perp$ is then aligned with the gravitational
field $\bm{g}$, namely
\begin{equation}\label{geq}
	\bm{g} \simeq \hat{\bm{\Pi}}_\perp \calW' \, .
\end{equation}
On the other hand, because $\sigma \ll \rho_\text{b}$ by the same mechanism,
we observe that the gravitational field equation \eqref{Poisson} with
\eqref{rhosigma} becomes
\begin{equation}\label{Poisson2}
	\bm{\nabla} \! \cdot \! \left( \bm{g} - 4 \pi G \, \bm{\Pi}_\perp
	\right) \simeq - 4 \pi G \, \rho_\text{b} \, ,
\end{equation}
which according to \eqref{Poisson_MOND} is found to be rigorously equivalent
to the MOND equation. Finally, by inserting the expression of the potential
\eqref{W} into \eqref{geq} and comparing with the defining equation
\eqref{Pig}, we readily find the MOND behaviour of the gravitational
susceptibility coefficient as
\begin{equation}\label{chiMOND}
	\chi(g) = - 1 + \frac{g}{a_0} + \mathcal{O}(g^2) \, ,
\end{equation}
in complete agreement with \eqref{recipe2}. We can thus state that the dipolar
fluid described by the action \eqref{S}--\eqref{L} explains the phenomenology
of MOND in a typical galaxy. Note also that in this model there is no problem
with the light deflection by galaxy clusters. Indeed the standard
general relativistic coupling to gravity implies the usual formula for the
bending of light.

To conclude, the present model reconciles in some sense the observations of
dark matter on cosmological scales, where the evidence is for cold dark
matter, and on galactic scales, which is the realm of MOND. In addition, it
offers a nice unification between the dark energy in the form of $\Lambda$ and
the dark matter in galactic halos. More work should be done to test the model,
either by studying second-order perturbations in cosmology, or by computing
numerically the non-linear growth of perturbations and comparing with
large-scale structures, or by studying the intermediate scale of clusters of
galaxies.


\end{document}